Bilinear log $n$ - log $p$ relation and critical power-law grain size distribution of crushable aggregates under compression and shear

Authors: Kan Sato *, Hiroko Kitajima **, Miki Takahashi ***, Takashi Matsushima *
* University of Tsukuba, Japan.
** Texas A&M University, USA.
***National Institute of Advanced Industrial Science and Technology, Japan.

Abstract:
In order to investigate the relation between the bulk plastic compression behavior and the evolution of grain size distribution (GSD) due to grain crushing under high-pressure compression and shear, we performed three types of loading experiments; single grain crushing (SGC) test, one-dimensional compression (ODC) test and rotary shear (RS) tests. The materials used are an angular mountain silica sand and a round river silica sand. The major findings are summarized as follows: (1) The SGC tests reveal that the Weibull model is successfully applied with the modulus $m$=2 for single grain crushing stress. (2) In the ODC tests, the relation between the applied pressure, $p$, and the resulting porosity, $n$, fits better on a bi-linear model in a log $n$ – log $p$ plot than in the classical $e$-log $p$ plot, where $e$ is the void ratio. (3) Both in the ODC and the RS tests, the GSD converges into a power-law (fractal) distribution with the exponent (fractal dimension) of about -2.5, which is close to the one for Apollonian sphere packing, -2.47 (Borkovec et al., 1994). (4) The proposed recursive pore filling model successfully describes the log $n$ – log $p$ relation in the ODC test and log $n$ – log $\gamma$ relation, where $\gamma$ is the shear strain, in the RS test in a consistent manner.

1. Introduction
Grain crushing is a phenomenon commonly observed in brittle granular materials under high pressure, and the resulting evolution of grain size distribution (GSD) drastically affects the evolution of their bulk mechanical properties including elasticity, shear resistance, permeability and so on. Understanding and modeling this behavior is quite important in various science and engineering problems such as bearing capacity of piles (Terzaghi et al. 1996), crushing process of rocks and concrete (Gudmundsson 2011, Matsushima et al. 2009), seismic fault slip (Lockner et al. 1991, Togo & Shimamoto 2012), and meteoroid impact on planetary surfaces (Melosh 1989).

Previous studies have clarified the following useful knowledges as well as remaining uncertainties. First, single grain crushing (SGC) experiments (Billam 1971, Lee 1992, Nakata et al. 1999) revealed the inverse correlation between the SGC stress and its grain size. Generally accepted explanation of this correlation is that the intra-grain defects are uniformly distributed and their strength distribution is described by the Weibull distribution (Weibull 1951, McDowell and Bolton 1998, Nakata et al. 1999,

Zhang et al. 2015, Stefanou and Sulem 2016). Among them, the recent work by Zhang et al. (2015) discussed the Weibull modulus observed in the experiments in terms of the intergranular contact mechanics.

Next, the stress-strain response of one dimensional compression test of crushable granular materials has often been modeled with bi-linear elasto-plastic model in $e - \log p$ plot, where $e$ is the void ratio and $p$ is the applied pressure (McDowell & Bolton 1998, Nakata et al. 2001). This $e - \log p$ bi-linear model is originated from the plastic consolidation model of clay (Schofield and Wroth 1968). Although this model has been very well established, it cannot be applied to very high pressure region because the corresponding void ratio cannot be negative but converge into zero. An alternative model to avoid this flaw is the $\log e - \log p$ bi-linear model (Pestana and Whittle 1995), but it is required to explore micromechanical understanding to establish the correct model. In particular, the relation between the SGC properties and this bulk compression behavior of crushable grains is a big issue to be solved.

Third, various observation so far proved that many of GSDs of the materials subjected to sufficient crushing process converge into power-law (or fractal) distribution (Jaeger 1967, Hartmann 1969, Kendall 1978, Turcotte 1986, Steacy & Sammis 1991, McDowell & Bolton 1998, Marks & Einav 2015). Above all, Turcotte (1986) summarized the power-law exponents (or the fractal dimensions) reported by many researchers for various fragmented objects that varies from 1.44 to 3.54. Although this fractal nature itself is understood by the absence of characteristic length both in crushing process and in material properties (Mandelbrot 1982), there is no established physical model to quantify this exponent. Recently Marks & Einav (2015) proposed a simple model to derive the exponent considering grain crushing, segregation and mixing. That kind of physics behind the grain crushing phenomenon should be pursued both from experimental and numerical sides.

Considering the circumstances described above, we have been investigating experimentally the stress-strain relation of crushable granular materials during high-pressure compression and shear, focusing on the evolution of GSD. In this particular paper, we firstly present the result of basic SGC tests of the two types of sand, an angular mountain sand and a round river sand, to discuss about the distribution of the SGC stress. Then, we show the results obtained in a series of one-dimensional compression (ODC) tests and rotary shear (RS) tests. Based on the experimental findings, we propose a simple recursive pore filling model, and show that the model can describe the experimental observation in a consistent manner.

2. Single grain crushing (SGC) test

We used two geo-materials in this study. One is Gifu sand, an angular mountain silica sand whose mean grain size is 2.38(mm). The other is Kashima sand, a round river sand whose mean grain size is 2.08(mm). The grain size of materials are carefully chosen so that (1) the SGC test can be performed easily and appropriately, (2) the resulting power law GSD can be examined in wider grain size range, (3) the ODC and the RS tests are performed with sufficient number of grains and with sufficient applied stress magnitude. Also, the two materials can be compared in the context of grain shape and natural process that the grains have been subjected to in the past.

The photos of the grains are shown in **Figure 1** and their mineral composition is listed in **Table 1**. According to the visual inspection, mountain sand grains can be categorized by two; transparent grains of quartz and non-transparent white grains of feldspar (**Figure 1(a)**). The number ratio of the feldspar grains is about 7.3% (151/(1924+151)=0.073). On the other hand, river sand consists of the grains of various colors (**Figure 1(b)**).

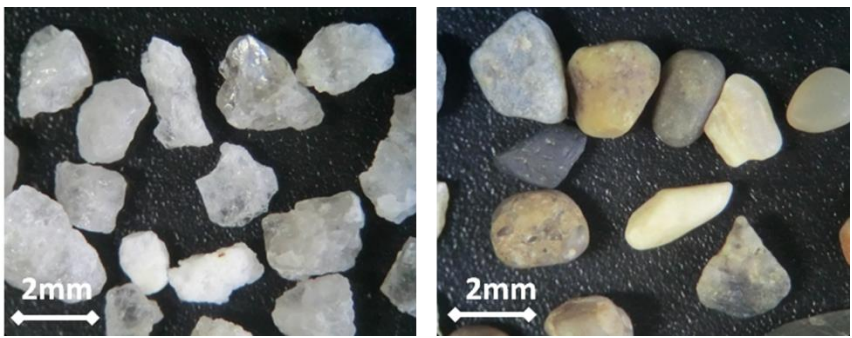

(a) Angular mountain silica sand (Gifu sand)     (b) Round river silica sand (Kashima sand)

**Figure 1**   Geo-materials used in the present study

**Table 1**   Mineral composition of the materials

|  | Mountain sand (Gifu) | River sand (Kashima) |
|---|---|---|
| Specific gravity | 2.59 | 2.64 |
| Diameter range (mm) | 0.420 - 2.360 | 0.850 - 3.350 |
| Chemical composition |  |  |
| $SiO_2$ (%) | 96.25 | 90.20 |
| $Al_2O_3$ (%) | 2.05 | 5.58 |
| $Fe_2O_3$ (%) | 0.04 | 0.85 |
| CaO (%) | 0.02 | 0.56 |
| MgO (%) | 0 | 0.2 |

The loading apparatus of the SGC test is shown in **Figure 2**. The loading speed is set to 0.015 (mm/s), being regarded as quasi-static loading. **Figure 3** shows some examples of the grains before and after the SGC test, and **Figure 4** shows their vertical load-displacement curves. The occurrence of the grain crushing was determined by the sudden drop of the load together with a visual inspection. Following

the previous studies (e.g., Nakata et al. 1999) the SGC stress, $\sigma_{SGC}$, is defined as:

$$\sigma_{SGC} = \frac{F_C}{h^2} \quad (1)$$

where $h$ is the height of the grains in the initial setup.

42 grains and 35 grains were tested for the mountain and the river sands, respectively, and the resulting distributions of $\sigma_{SGC}$ are shown in **Figure 5**. Firstly, it was observed that the river sand grains have much higher $\sigma_{SGC}$ than the mountain sand grains. This difference may come partly from the difference of grain shape; the river sand has relatively round grain shape that causes less stress concentration at the loading points than angular grains, and thus tends to exhibit higher $\sigma_{SGC}$. On the other hand, the difference of the natural process that the grains have experienced in the past may also be responsible for that; the river sand grains experienced a number of impact loading during the river transportation process and eventually the surviving grains have larger $\sigma_{SGC}$.

Secondly, more detailed examination for the mountain sand revealed that the feldspar grains have noticeably smaller $\sigma_{SGC}$ than the quartz grains. On the other hand, $\sigma_{SGC}$ of river sand grains does not differ in different grain color. This observation can also be understood by their origin; the mountain sand grains experienced less natural sorting process and accordingly keep the dispersity of $\sigma_{SGC}$ due to inherent rock strength than the river sand. On the other hand, the river sand consists of various rocks, but they have similar $\sigma_{SGC}$ by the natural mechanical sorting process of the river. Note that the number ratio of the feldspar grains is about 7.3%, and hence the collective behavior of the mountain sand may be governed by the property of the stronger quartz grains.

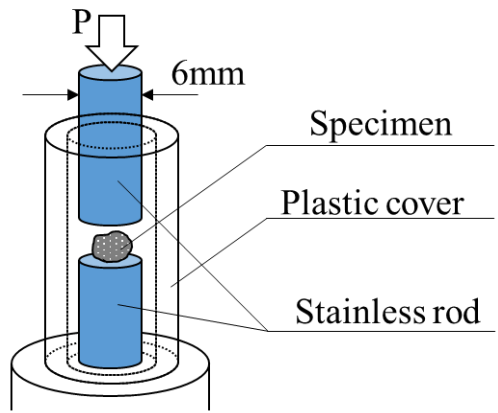

**Figure 2** Single Grain Crushing (SGC) test apparatus

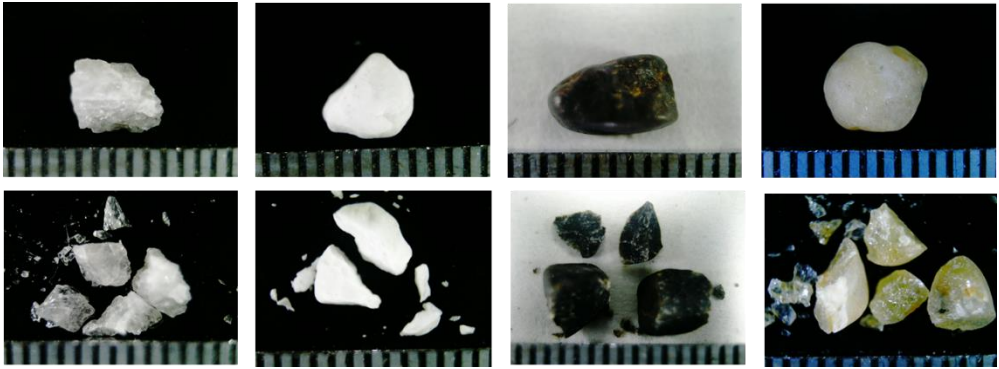

**Figure 3** Examples of the grains before and after single grain crushing test

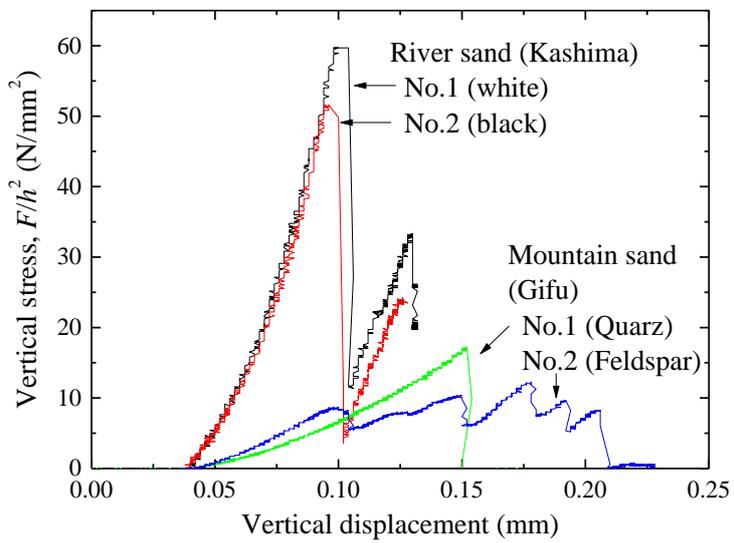

**Figure 4** Vertical load - displacement curves of SGC test for the grains shown in **Figure 3**.

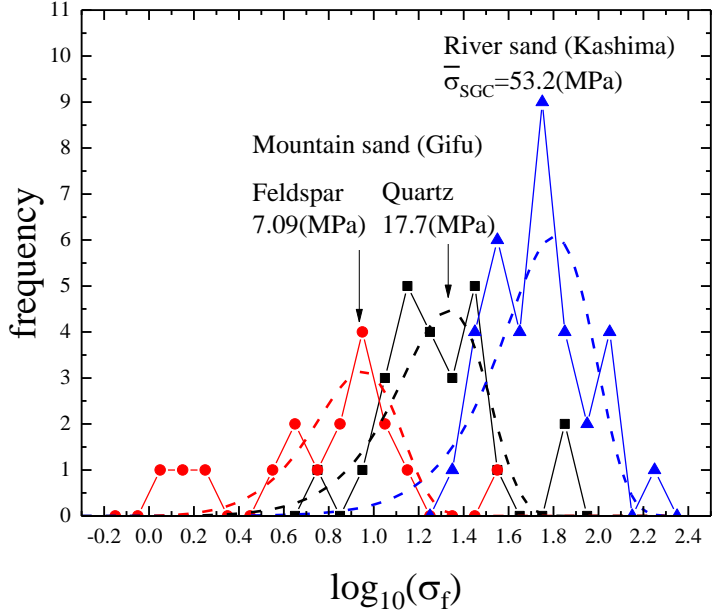

**Figure 5** Distribution of single grain crushing stress. The dashed lines are the fitted Weibull distribution.

Following the previous research (e.g., Nakata et al. 1999), the observed distribution of $\sigma_{SGC}$ was modeled by the Weibull distribution, in which the survival probability of grains is described by:

$$P_s(\sigma_{SGC}, V_G) = \exp\left[-\frac{V_G}{V_{G0}}\left(\frac{\sigma_{SGC}}{\sigma_0}\right)^m\right] \tag{2}$$

where $V_G$ is the grain volume, and $V_{G0}$ and $\sigma_0$ are the normalization parameters. Since the grain size range of the materials is not very wide, we assume that the effect of grain volume is negligible, and tried to fit the model into the SGC test results with $V_G/V_{G0}=1$, which is shown in **Figure 6**. It was found that Weibull model can describe the distribution of $\sigma_{SGC}$ fairly well with the Weibull modulus of about 2 both for the river sand grains, and two types of the mountain sand grains, respectively.

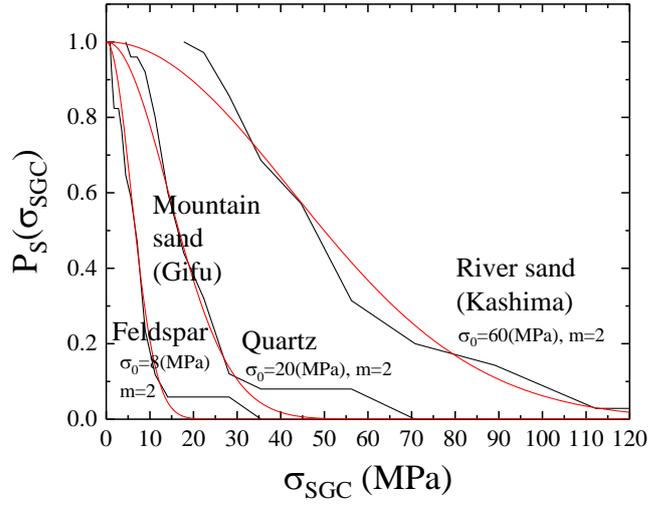

**Figure 6**   Survival probability of grains in SGC test and the fitted Weibull distribution

Once the Weibull modulus is determined, the average crushing stress is then calculated as a function of grain volume $V_G$ or the grain diameter $d$ as follows:

$$\bar{\sigma}_{SCG}(V_G) = \int_0^\infty \frac{dP_S}{d\sigma_{SGC}} \sigma_{SGC} d\sigma_{SGC} = \int_0^\infty \frac{dP_S}{dt} \sigma_{SGC} dt = \sigma_0 \left(\frac{d}{d_0}\right)^{-3/m} \Gamma(1+\frac{1}{m}) \qquad (3)$$

where $d_0$ is the normalization parameter and $\Gamma$ is the gamma distribution. The stress values shown in **Figure 5** are those obtained with this equation for each type of sand grain.
This expression will be used in Section 5.

3. One-dimensional compression (ODC) test

Our one-dimensional compression device is composed of a stainless (SUJ2) cell with its inner diameter of 30.0(mm), the rod of the same material and the loading frame (**Figure 7**). The nominal yield stress of SUJ2 is 1370 (MPa). The capacity of the loading frame is 200 (t). The loading speed is 10 (μm/s) for all the cases. Since the specimen height is about 20 (mm), the compressive strain rate is about $5\times10^{-4}$ (1/s), which can be regarded as a quasi-static loading.

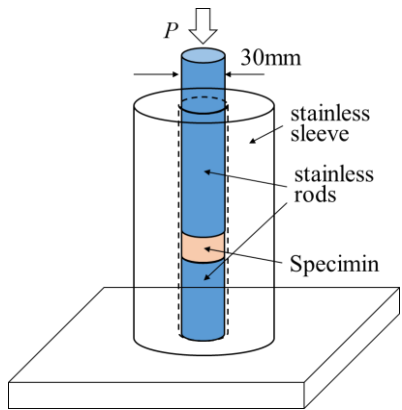

**Figure 7** One dimensional compression (ODC) test setup

The materials used in the experiment are the same as those in the previous section. Each specimen was prepared in a loose condition using a funnel, and a vertical loading was applied up to various level (140 (MPa) to 700 (MPa)). The measured volumetric strain was modified considering the deformation of stainless rods, sleeve and load cell. **Figures 8(a)** and **8(b)** shows the relation between the void ratio, $e$, and the vertical loading pressure, $p$, of the tests for the mountain sand in two different ways: the classical $e$-log $p$ plot and the log $e$-log $p$ plot proposed by Pestana and Whittle (1995). It is obvious from the figures that the classical $e$-log $p$ bilinear model cannot be applied in a very high pressure region, while the log $e$-log $p$ plot seems to be described by a bilinear model. Considering the fact that the void ratio cannot be negative even under very high pressure, the log $e$-log $p$ bilinear model is a reasonable model. We will discuss this issue further in the following sections.

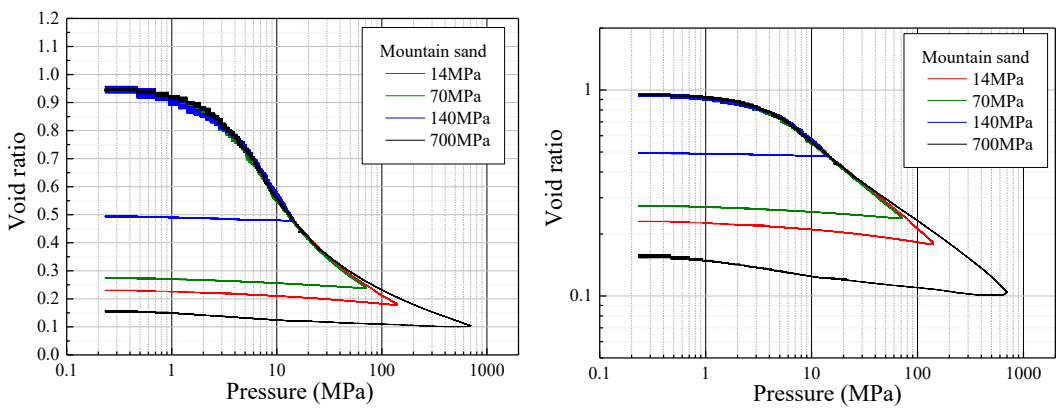

**Figure 8** Relation between the applied pressure and the resulting void ratio of the specimens. $e - \log p$ plot (left) and $\log e - \log p$ plot (right) for the mountain (Gifu) sand

**Figure 9** is the comparison between the mountain (Gifu) sand and the river (Kashima) sand in $\log e - \log p$ plot. The difference of the initial void ratio between the two materials (0.957 and 0.723 in the mountain and the river sands, respectively) is caused by the difference of their grain shape. The yield stress $\sigma_Y$ (which is regarded as the pressure where the slope of the curve suddenly changes as marked in the figure) is about 4(MPa) for the mountain sand, and about 15(MPa) for the river sand. The ratio of $\sigma_Y$ to $\bar{\sigma}_{SGC}$ for each sand is about 0.25 to 0.3, which is consistent with the previous research (Nakata et al. 2001). The difference between $\sigma_Y$ and $\sigma_{SGC}$ is caused by the heterogeneous stress transmission in granular medium, which is known as "force chain" (Oda and Konishi 1974, Majmudar and Behringer 2005). This issue is also discussed in details in the next section.

**Figure 9** also shows that the power-law exponent in the grain crushing regime is almost identical for the two materials. The value is about -0.38 to -0.40.

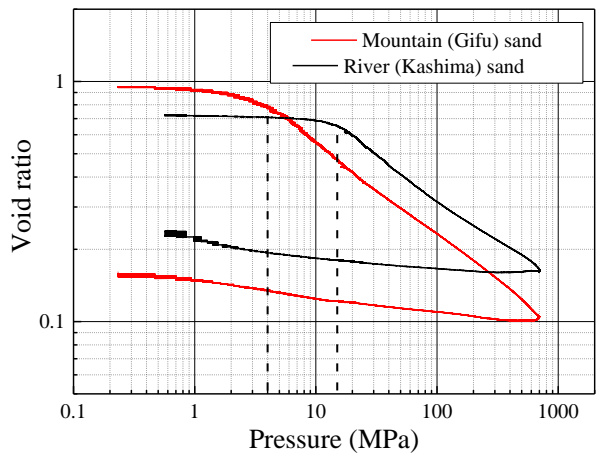

**Figure 9** Comparison between the mountain sand (Gifu sand) and the river sand (Kashima sand) in $\log e - \log p$ plot

**Figure 10** shows the materials after loading up to 700 (MPa). It can be noticed that some grains remain uncrushed by the cushioning effect (e.g., Marks & Einav 2015). The grain size distribution (GSD) after loading was measured both by an optical particle analyzer (Camsizer P4) and by a laser diffraction particle size analyzer (Seishin Enterprise, LMS-3000e). Since the former is applicable to the grains larger than 33(μm) while the latter can measure the grains whose size is between 0.1(μm) and 3,500(μm), we sieved the original material to make two samples suitable for each apparatus, and connected the two results considering the best matching in the overlapped range (33(μm)<D<3,500(μm)).

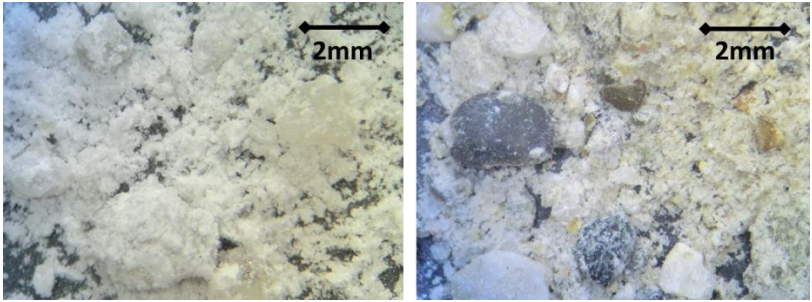

(a) Angular mountain silica sand (Gifu sand)     (b) Round river silica sand (Kashima sand)

**Figure 10** Microscope photos of the materials after loading up to 700 (MPa). Compared with **Figure 1**, it can be recognized that some original grains remain uncrushed.

The resulting GSDs of the mountain sand for various loading levels are shown in **Figure 11** in volume cumulative. As a general trend, the GSDs are shifted to the left in increasing loading level due to the grain crushing. However, if one checks the GSD curves carefully, it is noticeable that they have some local slope changes. In order to make clear this tendency, the derivative of those curves, which is the volume frequency, is shown in **Figure 12**. The figure clearly shows the second and the third peaks appearing in the progress of grain crushing. Note that such local peaks are also seen in Fig. 7 in Nakata et al. (2001). One can roughly estimate from **Figure 12** that the ratio of the second peak grain size to the first one (the original grain size) is about 0.15, and the ratio of the third one to the second is about 0.05. This may correspond to the representative pore size of the granular packing structure, which will be discussed in the Section 5.

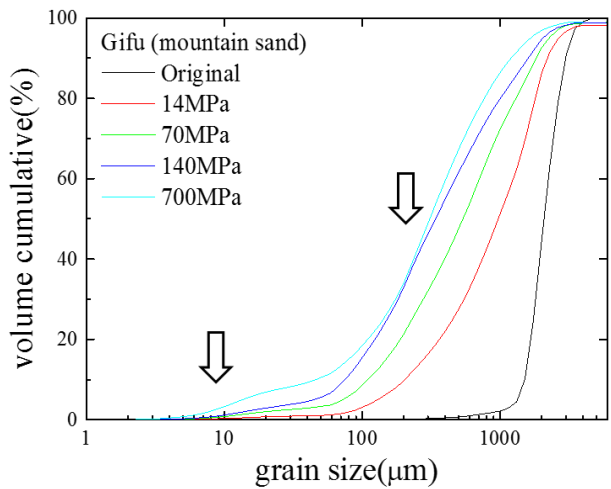

**Figure 11** Grain size distribution (GSD) for different loading levels shown by volume cumulative (Gifu mountain sand). Arrows show some local maxima of the slopes of the curves for 140(MPa) and 700(MPa).

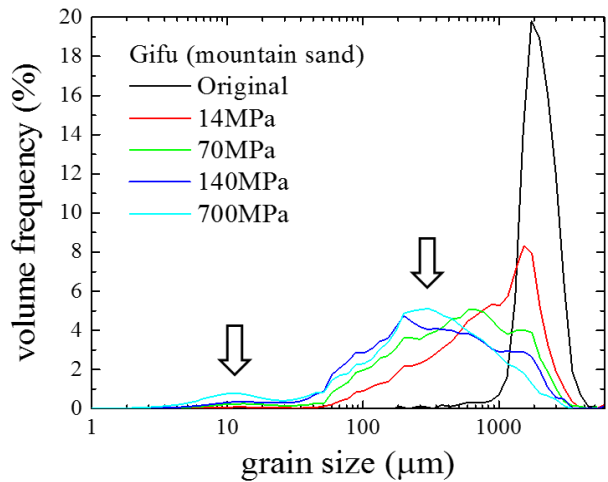

**Figure 12** GSD for different loading levels shown by volume frequency (Gifu mountain sand). Arrows show the same local maxima shown in **Figure 11**.

As described in Introduction, the GSD after crushing process has been discussed in the context of the power-law (or fractal) distribution in the literature (Hartmann 1969, Turcotte 1986, Steacy and Sammis 1991, McDowell & Bolton 1998), where the cumulative number of grains whose diameter is greater than $d$ was plotted against $\log d$. This kind of GSD plot for our results is shown in **Figures 13** and **14**. **Figure 13** is the GSDs for different loading levels of the mountain sand, while **Figure 14** shows the comparison between the mountain sand and the river sand. According to the figures, the GSDs after the sufficient progress of grain crushing have the power-law nature with its exponent (or the fractal dimension, the slope of the curve, denoted by $D_f$) of about -2.5 over the three order of magnitude. The similar power-law exponent was observed by Matsushima et al. (2014) for Toyoura sand as well.

It is quite instructive that this power-law exponent is close to that of Apollonian sphere packing, -2.47 (Borkovec et al. 1994). In this packing protocol, starting from a mono-dispersed random spheres pack, the voids are filled with the largest possible grains. This process is repeated and finally some fractal structure is obtained. The similarity of the exponent between this pack and the ODC tests implies that the grain crushing under such confined condition as ODC tests is governed by the similar "pore filling" mechanism.

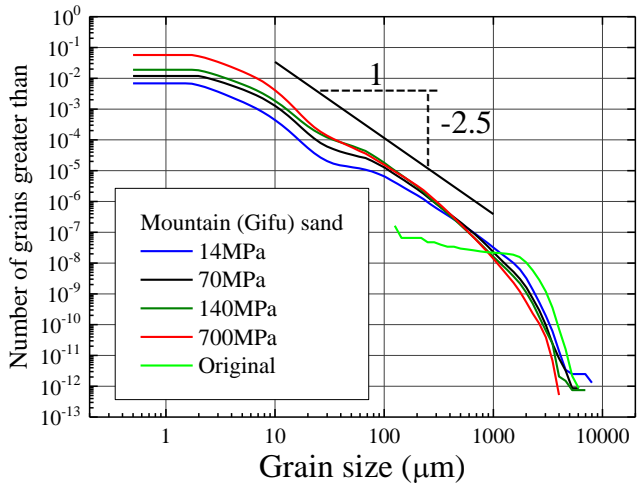

**Figure 13** GSDs for different loading levels for the mountain sand by the cumulative number

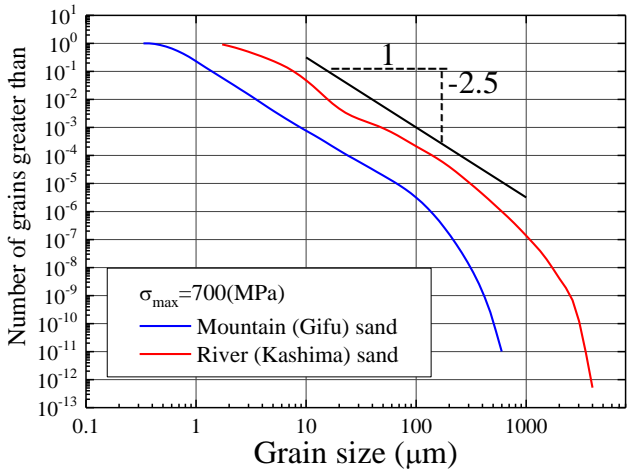

**Figure 14** Comparison of the final GSDs between the mountain and the river sand by the cumulative number

4. Rotary Shear (RS) test

The apparatus used in the RS tests has been developed by Kitajima et al. (2010). In the apparatus, an air actuator applies the constant axial load (normal stress) at one end of a sample via a granite rock cylinder, while an electric motor drives rotation at the other end via a cylinder of the same rock (**Figure 15**). The sample diameter is 25 (mm) and its initial height before rotational shear is about 3 to 3.5 (mm). The samples were prepared in a loose condition, whose measured void ratio was ranging from 0.9 to 1.1.

In the present study, the rotation rate was set to 0.75, 75, and 750 (rpm), whose corresponding shear rates $\dot{\gamma}$ are 0.21, 21, and 210 (1/s), respectively, at the place of the volume equivalent radius, $25 \times 2/3$ (mm). All samples were subjected up to 100 revolutions that reaches about $1.7 \times 10^3$ shear strain in average. The normal stress was kept constant ($\sigma_n = 1.0$ (MPa)) during the shear.

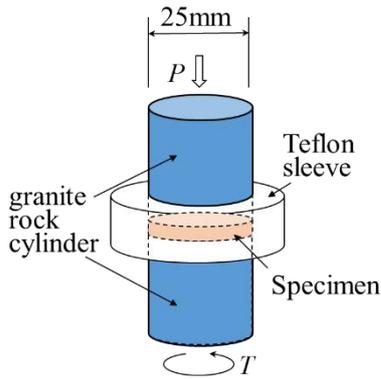

**Figure 15**  Rotary shear (RS) apparatus

**Figure 16** shows the evolution of shear stress ratio, $\tau/\sigma_n$, in terms of shear strain $\gamma$ for the two sands with three different shear rates. According to the figure, $\tau/\sigma_n$ developed up to about 0.6 in the early stage of the shear, and kept almost constant until the end. This tendency is more or less common for all tests.

The corresponding void ratio evolution is shown in **Figure 17**. Compared with **Figure 16**, it is clear that the rapid decrease of the void ratio to around 0.1 to 0.2 is accompanied by the increase of shear stress. **Figure 18** shows this correlation more clearly; it is almost linear throughout the entire shear process. The reason of this correlation is not very clear, but considering that $\tau/\sigma_n=0.3$ at $e=1.0$ is rather small for the shear strength of usual sands, there is a possibility that the slip occurs between the sand and the boundary rock cylinder.

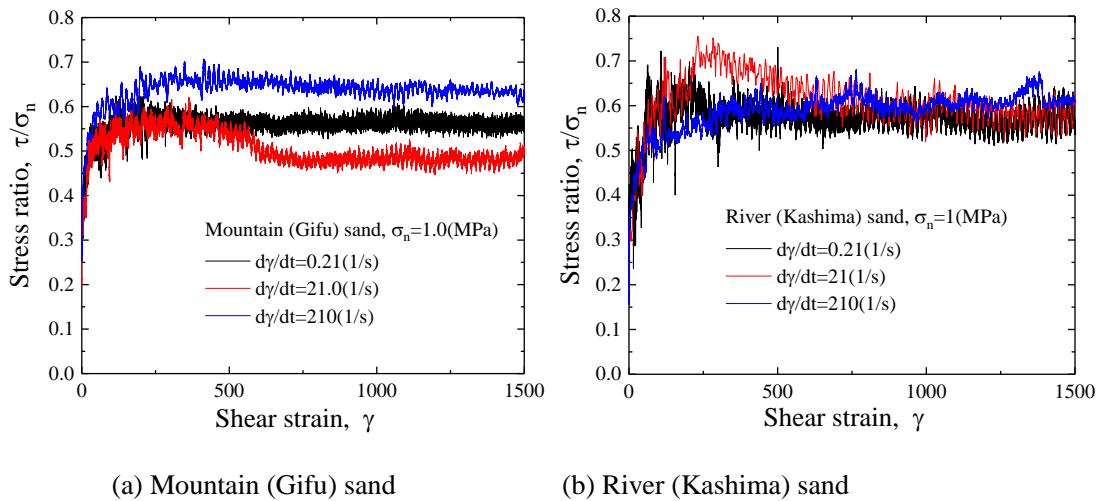

(a) Mountain (Gifu) sand     (b) River (Kashima) sand

**Figure 16** Shear stress - shear strain curve for three different shear rates.

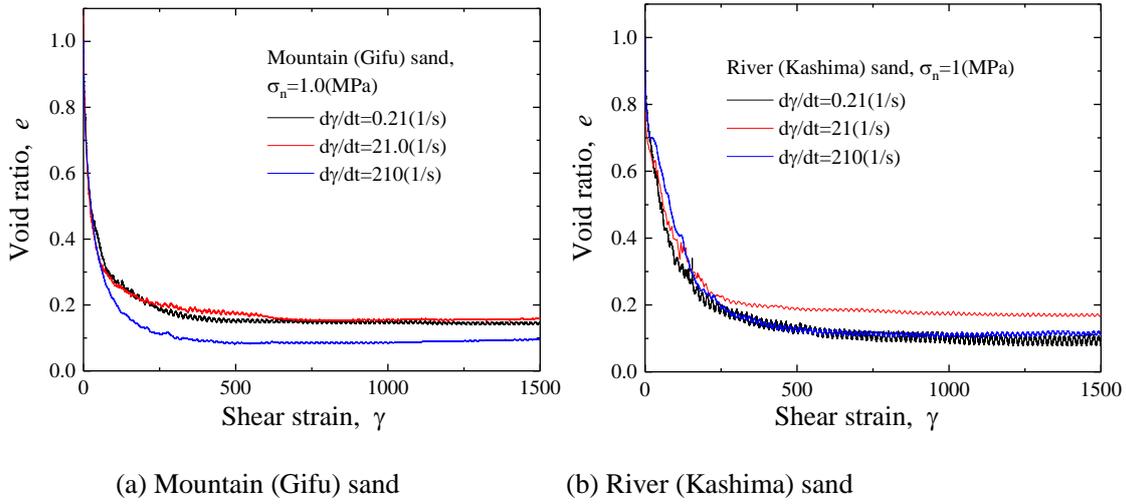

(a) Mountain (Gifu) sand  (b) River (Kashima) sand

**Figure 17** void ratio - shear strain curve for three different shear rates.

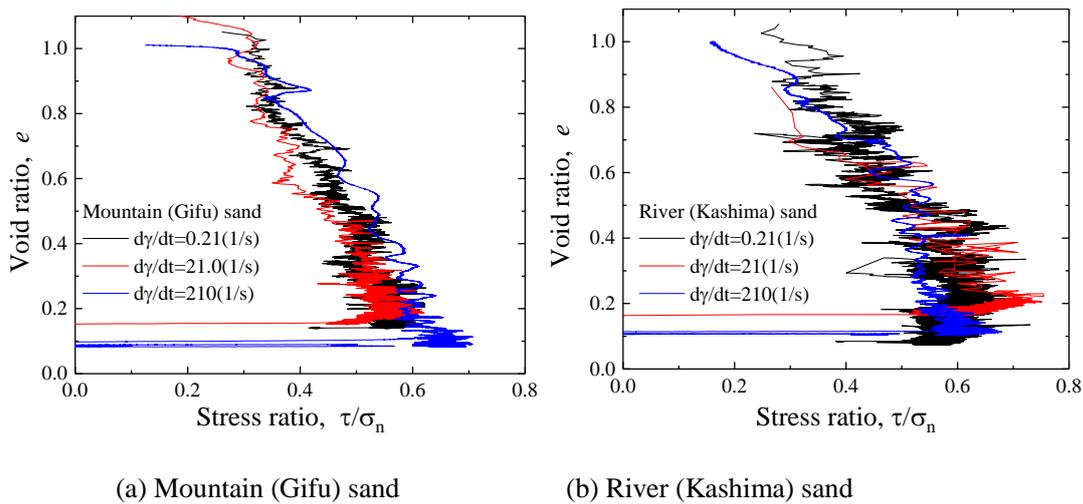

(a) Mountain (Gifu) sand  (b) River (Kashima) sand

**Figure 18**  Relation between void ratio and stress ratio for three different shear rates.

It should be noted that the grain crushing seems to occur with much smaller stress level than the yield stress $\sigma_Y$ in the ODC tests (see **Figure 9**). This can be understood by continuous rearrangement of grain structure in the RS tests. During this shearing process, some grains that happen to have very large contact force crush. On the other hand, the ODC test has an opportunity of grain structure rearrangement only by grain crushing event, and a weak granular structure changes into the stronger one by pore-filling of fine powders.

Despite such difference between the ODC test and the RS test, the final GSD in the RS test, shown in **Figures 19** and **20**, are surprisingly similar to those in the ODC test; the second and the third peaks of the GSDs are observed in all cases, and the power-law exponent of GSD is close to -2.5 over three order of magnitude. Shear rate does not seem to affect the final GSD. These features clearly indicate the importance of the pore filling mechanism in grain crushing process in the RS tests as well.

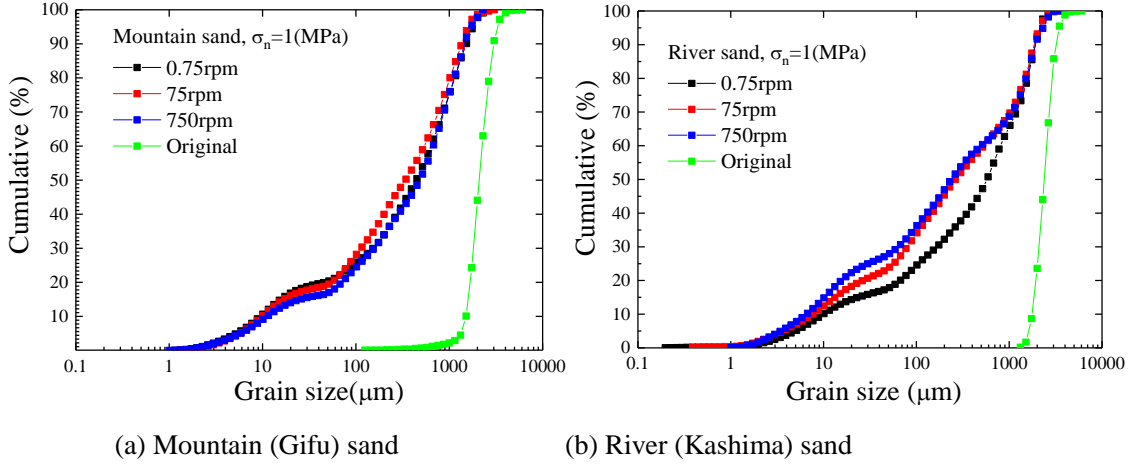

(a) Mountain (Gifu) sand  (b) River (Kashima) sand

**Figure 19** Final GSDs for three different shear rates

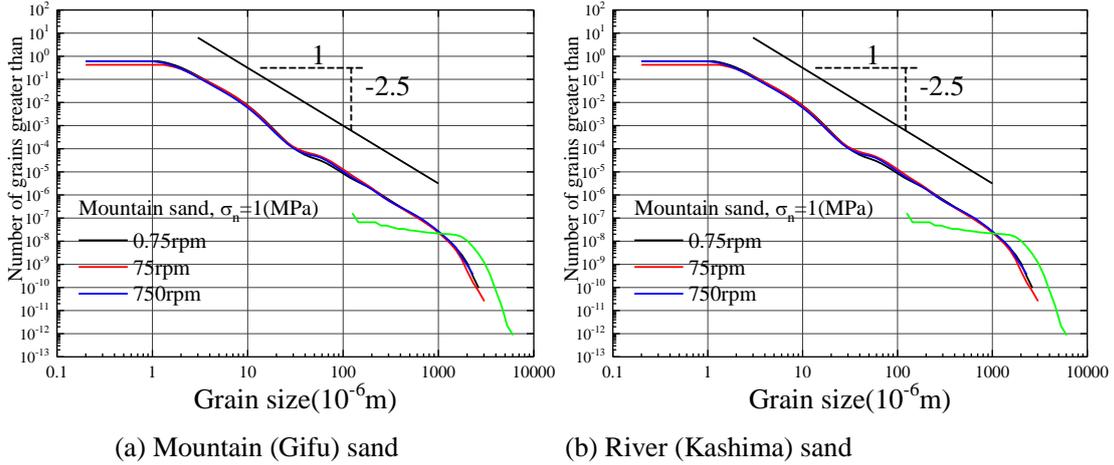

(a) Mountain (Gifu) sand  (b) River (Kashima) sand

**Figure 20** Final GSDs for three different shear rates by the cumulative number

## 5. Model

### 5.1 Recursive pore filling model

Based on the aforementioned experimental results, we construct a model which connects the bulk plastic compression behavior of crushable aggregates with the evolution of GSD. First, we assume an assembly of mono-disperse grains as an initial state. The void ratio of the initial assembly $e_0$ is described by:

$$e_0 = \frac{V_{void}}{V_{solid}} \qquad (4)$$

where $V_{solid}$ is the total volume of the initial grains, and $V_{void}$ is the total volume of the void.

Then we consider to fill the void $V_{void}$ with the smaller grains with the same void ratio, $e_0$, that is:

$$e_0 = \frac{V_{void} - V_{solid}^S}{V_{solid}^S} \tag{5}$$

where $V_{solid}^S$ is the total volume of the smaller grains. Now the number ratio of the smaller grains to the larger grains is:

$$\frac{N_S}{N_L} = \frac{V_{solid}^S}{V_{solid}}\left(\frac{d_L}{d_S}\right)^3 = \frac{e_0}{1+e_0}\alpha^{-3} \tag{6}$$

where $\alpha \equiv (d_S/d_L)$ is the size ratio of the smaller grain to the larger initial grain. Equation (6) can be rewritten as:

$$\log N_S - \log N_L = \log\left(\frac{e_0}{e_0+1}\right) - 3(\log d_S - \log d_L) \tag{7}$$

and thus we can define a discrete version of the fractal dimension (power-law exponent) of the GSD as:

$$D_f \equiv -\frac{\log N_S - \log N_L}{\log d_S - \log d_L} = 3 - \frac{1}{\log \alpha}\log\left(\frac{e_0}{e_0+1}\right) \tag{8}$$

**Figure 21** shows the relation between $D_f$ and $\alpha$ for different $e_0$, computed from Equation (8). If we assume that $D_f$ is about -2.5, the value observed both in the ODC tests and in RS tests, we found that the corresponding $\alpha$ is about 0.14 to 0.25 for usual void ratio range of poorly-graded aggregates ($0.6 < e_0 < 1.0$) in the figure. This value agrees with the local maxima of the GSD observed in the tests. Therefore it can be said that Equation (8) describes well the relation between $D_f$ and $\alpha$ during those crushing process.

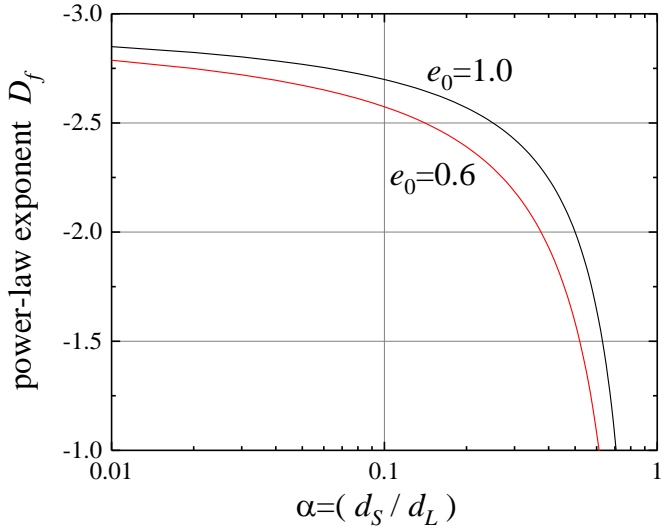

**Figure 21** Relation between the fractal dimension (power-law exponent) of GSD and the size ratio of the smaller grain to larger grain in Equation (8).

Next, the overall void ratio after adding the small grains, $e_1$, satisfies the following relation:

$$e_1 = \frac{e_0}{\frac{e_0+1}{e_0}+1} \qquad (9)$$

This is regarded as a recursive equation, and the void ratio after repeating this process $k$ times is:

$$1+\frac{1}{e_k} = \left(1+\frac{1}{e_0}\right)^{k+1} \quad \text{or} \quad n_k = n_0^{k+1} \qquad (10)$$

where $n \equiv V_{void}/(V_{solid}+V_{void}) = e/(1+e)$ is the porosity. Since the radius of the smallest grains is $d_k = \alpha^k d_0$, we have:

$$k = \frac{\log \frac{d_k}{d_0}}{\log \alpha} \qquad (11)$$

Equations (10) and (11) yield the relation between $e_k$ and $d_k$ or $n_k$ and $d_k$ as:

$$\frac{\log(1+\frac{1}{e_k})}{\log(1+\frac{1}{e_0})} = \frac{\log n_k}{\log n_0} = k+1 = \frac{\log \frac{d_k}{d_0}}{\log \alpha}+1 \qquad (12)$$

*5.2 Application into ODC tests*

Following McDowell and Bolton (1998), we assume that the breakage stress for the smallest grain determines the ODC stress. Note that this assumption comes from the cushioning effect; a large grain surrounded by a lot of small grains tends to remain uncrushed because a lot of contact points with small grains create more isotropic stress field in the grain (e.g., Marks and Einav 2015). Together with the Weibull model described above, the minimum grain size of the crushable granular assembly, $d_{min}$, subjected to the compressive stress, $p$ is described as follows:

$$p = C \cdot d_{min}^{-3/m} \quad \text{or} \quad -\frac{3}{m}\log\left(\frac{d_{min}}{d}\right) = \log(p/\sigma_Y) \qquad (13)$$

Setting that $d_{min}$ equals to $d_k$ yields the final relation between the overall void ratio $e$ ($=e_k$) and $p$ as:

$$\frac{\log(1+\frac{1}{e_k})}{\log(1+\frac{1}{e_0})} = k+1 = \frac{\log\frac{d_k}{d}}{\log\alpha}+1 = \frac{-\frac{m}{3}\log(p/\sigma_Y)}{\log\alpha}+1 \qquad (14)$$

**Figure 22** shows the relation between $\log e_k$ and $\log(p/\sigma_Y)$ obtained from this equation. Note that $m=2$, which is assumed in the plot, is observed in the SGC test, and $(e_0, \alpha)=(1.0, 0.25)$ or $(e_0, \alpha)=(0.6, 0.14)$ are obtained in **Figure 21**. Not perfect but almost linear relation is observed, and the resulting exponent is about -0.36 to -0.40, which agrees well with the slope observed in the ODC test shown in **Figures 8** and **9**. In fact, the equation converges a linear relation between $\log e_k$ and $\log p$ when $e$ is approaching zero. The compression coefficient defined in $\log e$ - $\log p$ plot is then:

$$C_C \to \frac{m}{3\log\alpha}\log(1+\frac{1}{e_0}) \qquad \text{when } e_k \to 0 \qquad (15)$$

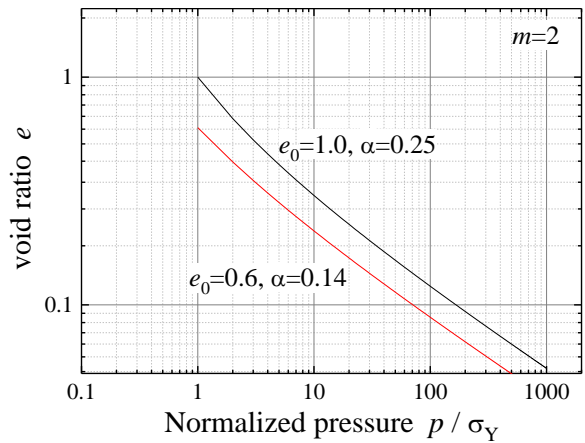

**Figure 22** log e - log p relation predicted by Equation (14)

Alternatively, using the porosity $n$, Equation 14 is rewritten as:

$$\frac{\log n_k}{\log n_0} = \frac{-\frac{m}{3}\log(p/\sigma_Y)}{\log \alpha} + 1 \qquad (16)$$

which shows a perfectly linear relation between $\log n$ and $\log p$. The validity of this equation is demonstrated in **Figure 23** in which the ODC test data show good straight lines for both sands and dashed lines are the relation computed by Equation (16) with best fit values of $\alpha$. These values are also consistent both with **Figure 21** and with the local maxima observed in **Figure 12**.

To authors' knowledge, this is the first success to explain the compression curve (usually plotted as $e - \log p$ curve) commonly observed in ODC tests of crushable sands and aggregates by their grain scale micromechanics.

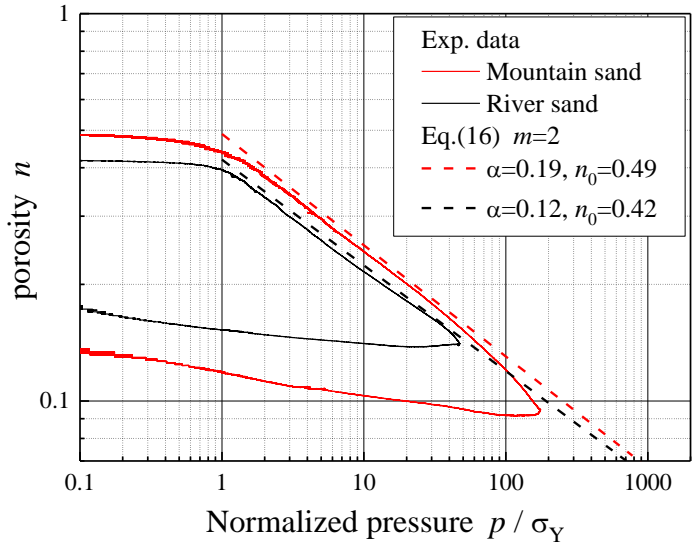

**Figure 23** $\log n$ - $\log p$ relation in ODC tests. Solid lines are the experimental data replacing the vertical axis $\log e$ by $\log n$ in **Figure 9**. Dashed lines are the best fit with Equation (16).

*5.3 Application into RS tests*

Here, we apply the aforementioned recursive pore-filling model to describe the shear crushing behavior observed in the RS tests. We assume again the step-by-step crushing process shown in Equation (10), and consider the crushing process of the k-th grains (whose size is $d_k$) into the (k+1)-th grains (whose size is $d_{k+1}$) during shear loading. Since the granular packing structure continuously changes during shear deformation, we assume that the structure is refreshed for every unit shear strain, $\hat{\gamma}_k$, and that the surviving grains in the current shear strain, denoted by $(i \times \hat{\gamma}_k)$ where $i$ is a shear strain step number, will newly crush with the probability $(1 - P_S)$, where $P_S$ is the survival probability of a single grain described by Equation (2). Based on this assumption, the number of the surviving k-th grains, $N_k$, and the number of the created (k+1)-th grains, $N_{k+1}$, after $(i \times \hat{\gamma}_k)$ strain

are described respectively by:

$$N_k = N_{k,0} \cdot (P_S)_k^{i} \qquad (17)$$

$$N_{k+1} = (1-(P_S)_k^{i})\frac{N_{k,0}}{\alpha^3} \qquad (18)$$

where $N_{k,0}$ is the number of the k-th grains in the beginning of the k-th process. Then, the condition that the (k+1)-th grains entirely fill the void of the k-th grains are described from Equation (6) as:

$$\frac{N_{k+1}}{N_k} = \frac{e_0}{1+e_0}\alpha^{-3} \qquad (19)$$

Equations (2), (17)-(19) yield:

$$i = \frac{\ln\left(\frac{1+e_0}{1+2e_0}\right)}{\ln(P_S)_k} = \frac{\ln\left(\frac{1+e_0}{1+2e_0}\right)}{-\alpha^{3k}\beta^m} \qquad (20)$$

where $\beta \equiv \sigma_p / \sigma_0$ is the ratio of the representative compressible stress in the sheared granular specimen, $\sigma_P$, to the normalized value $\sigma_0$. If the shear stress ratio $\tau/\sigma_n$ is assumed to be constant, $\beta$ is also constant and is proportional to $\sigma_n$.

This equation provides us the shear strain necessary for the k-th process, $(i \times \hat{\gamma}_k)$, and accordingly can be transformed into the following differential form:

$$\frac{d\gamma}{dk} = \frac{\ln\left(\frac{1+e_0}{1+2e_0}\right)}{-\alpha^{3k}\beta^m}\hat{\gamma}_k = \frac{\ln(1+n_0)}{\alpha^{3k}\beta^m}\hat{\gamma}_k \qquad (21)$$

where $n_0 = e_0/(1+e_0)$ is the initial porosity. Regarding the unit shear strain for the granular structural refreshment, $\hat{\gamma}_k$, it is obvious that the local strain in the pore space is much larger than the bulk shear strain because the grains are almost rigid (**Figure 24**), and its ratio may be related to the ratio of the void to the total volume. Therefore, we assume in the model that $\hat{\gamma}_k = \hat{\gamma}_0 \cdot n_0^{k}$ where $\hat{\gamma}_0$ is a constant value. Finally, solving this differential equation, we obtain the resulting total shear strain after the k-th process as:

$$\gamma = \frac{\hat{\gamma}_0}{\beta^m}\frac{\log(1+n_0)}{\log(n_0/\alpha^3)}\left[\left(\frac{n_0}{\alpha^3}\right)^k - 1\right] \equiv \Gamma\left[\left(\frac{n_0}{\alpha^3}\right)^k - 1\right] \qquad (22)$$

The above equation together with Equation (10) provides the relation between the applied shear strain and the resulting porosity as:

$$\frac{\log n}{\log n_0} = \frac{\log(\gamma/\Gamma + 1)}{\log(n_0/\alpha^3)} + 1 \qquad (23)$$

which indicates the linear relation between $\log n$ and $\log \gamma$ with the slope $\kappa$ of

$\log n_0 / \log(n_0 / \alpha^3)$ when $\gamma \gg \Gamma$.

Now we try to compare the model with the RS test results. If $n_0 = 0.5$ ($e_0 = 1.0$) is assumed, $\kappa = -0.1115$ (when $\alpha = 0.1$) and $\kappa = -0.1676$ (when $\alpha = 0.2$). The value of $\Gamma$ is rather difficult to estimate. First, the granular structural refreshment shear strain for the initial mono-disperse system, $\hat{\gamma}_0$, may be around 1.0, which corresponds to the displacement of each grain in the shear direction is identical to the grain size. Next, assuming $\sigma_P$ is comparable to the vertical stress $\sigma_n$, $\beta \equiv \sigma_p / \sigma_0$ ~1/60 in Kashima river sand and $\beta$ ~1/20 in Gifu mountain sand. $m$ is set to 2.0 based on **Figure 6**. Accordingly, considering the range of $\alpha$ ($0.1 < \alpha < 0.2$), $\Gamma$ ~230-350 for the river sand, and $\Gamma$ ~26-55 for the mountain sand.

**Figure 25** shows log $n$- log $\gamma$ curves calculated by Equation (23) using the parameters described above. It is clearly seen that $\Gamma$ represents the threshold shear strain from which the porosity begins to decreases linearly and that $\alpha$ affects the porosity decreasing rate, $\kappa$. On the other hand, **Figure 26** (a)(b) are the corresponding plots in RS tests for the mountain sand and the river sand, respectively. One can find that the basic trend is the same as the model responses in **Figure 25**, except that the porosity converges into the value of around 0.1 in the experiments. The existence of this critical state may be due to the increase of the crushing resistance of fine adhesive particles whose size is less than 1 (μm). The behavior of this regime is not incorporated in the present model.

More detailed comparison between **Figure 25** and **Figure 26** revealed that the threshold shear strain $\Gamma$ for the mountain sand and for the river sand roughly agrees with those in the models for $\Gamma = 30$ and 300, respectively, while $\kappa$ in the RS tests is larger than those calculated in the models. The reason of this discrepancy has not been clarified yet, but the heterogeneous shear strain in RS tests in radial direction could be responsible for that.

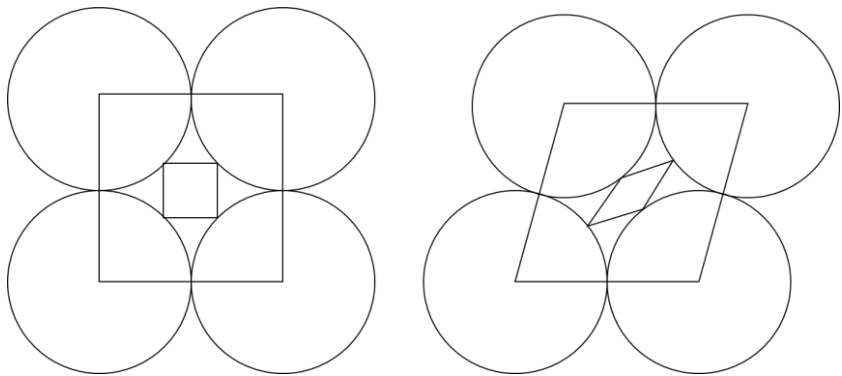

**Figure 24** A conceptual illustration showing that the local strain in the pore space is much larger than the bulk shear strain

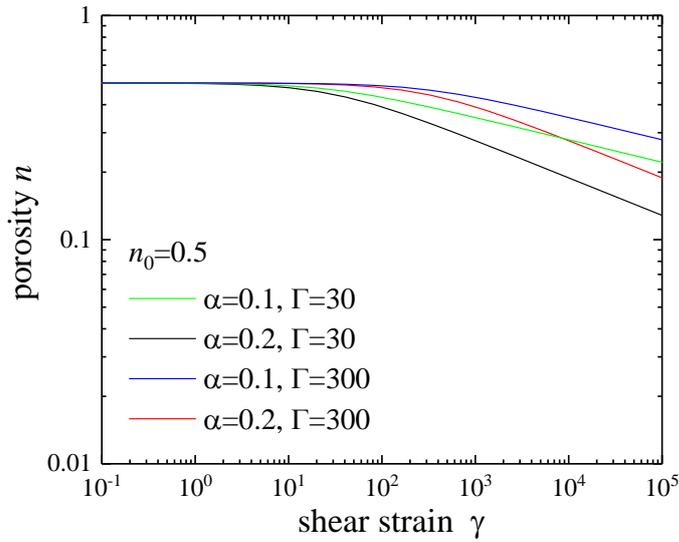

**Figure 25** Porosity change due to shear strain application calculated by Equation (23)

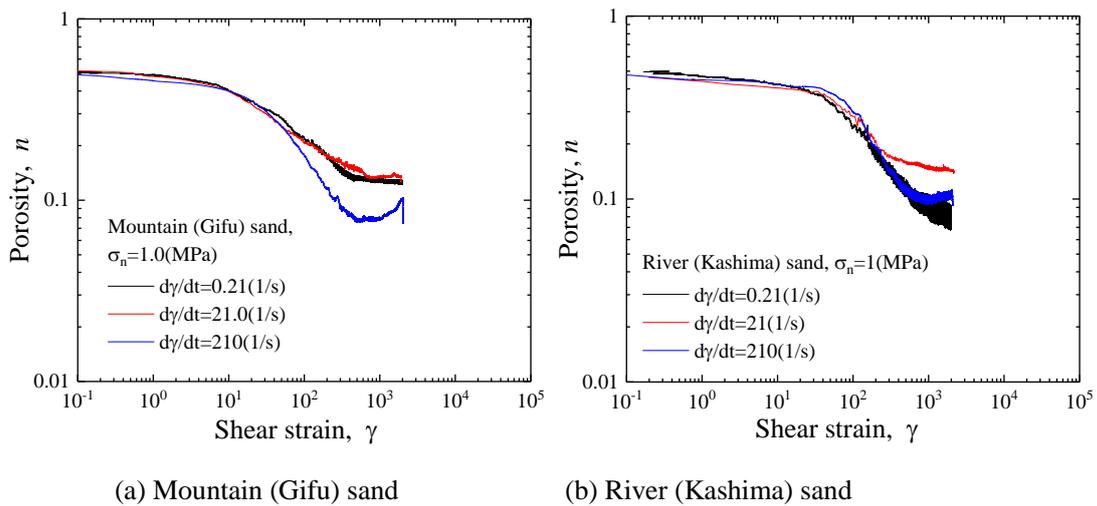

(a) Mountain (Gifu) sand  (b) River (Kashima) sand

**Figure 26** Porosity - shear strain curves of RS tests in log-log plot

6. Conclusions

Three types of loading experiments; single grain crushing (SGC) test, one-dimensional compression (ODC) test and rotary shear (RS) tests were performed for the angular mountain (Gifu) sand and a round river (Kashima) sand. The SGC tests revealed that the Weibull model is successfully applied with the modulus $m=2$ for single grain crushing stress and that the crushing stress of the river sand is about 3 to 6 times larger than that of the mountain sand. In the ODC tests, the yield stress clearly reflects the difference of grain crushing stress in SGC tests, and the relation between the applied pressure, $p$, and the resulting porosity, $n$, fits well on the bi-linear model in a log $n$ – log $p$ plot rather than in the classical $e$-log $p$ plot. In RS tests, grain crushing occurred in much lower stress level than ODC tests due to the extremely large shear strain application. Both in the ODC and the RS tests, the

GSD converged into the power-law (fractal) distribution with the exponent (fractal dimension) of about -2.5, which is close to the one for Apollonian sphere packing, -2.47 (Borkovec et al., 1994). Based on the above observation, we proposed the recursive pore filling model, and it was demonstrated that the model successfully describes the experimental data including the log *n* – log *p* relation in the ODC test and log *n* – log $\gamma$ relation in the RS test.

The most outstanding result in this paper is that the proposed model is based on the grain scale geometry and mechanics and can connect the bulk plastic compression behavior, commonly described by *e*-log *p* curve, with the evolution of grain size distribution (GSD) due to grain crushing. The quantitative consistency both for compression and shear loading implies that the model captures the correct physics of collective grain crushing behavior.

This paper deals only with the grain crushing under confined condition, but the model can apply to the behavior under unconfined condition. According to **Figure 21**, the GSD power-law exponent can be larger than -2.5 if the void ratio is larger than 1.0. This may happen under high speed shear such as impact cratering (Melosh 1989) for example, which may be the reason why the GSD power-law exponent of the lunar soil ranges from -3.0 to -4.0 (Tsuchiyama et al. 2011). In such a case, however, the grain size ratio $\alpha$ in the model must be determined by another mechanism like crack pattern geometry in a grain.

Another important extension of the present study is the modeling of cray consolidation behavior. If we deal with the collapse of cray meso-structure same as the crushing of a single solid grain (Suzuki and Matsushima 2012), it is quite natural to think that the cray consolidation can be described in a similar framework.

Acknowledgment

The authors acknowledges M. Iidaka of University of Tsukuba for manufacturing experimental equipment, M. Seta for performing preliminary experiments, and Prof. I. Einav of University of Sydney for valuable discussion. This work was supported by JSPS KAKENHI Grant Number JP26289150.

References

Terzaghi, K., Peck, R. B., & Mesri, G. (1996). Soil mechanics in engineering practice. John Wiley & Sons.

Gudmundsson, A. (2011). Rock fractures in geological processes. Cambridge University Press.

Matsushima, T., Ikema, T., & Yamada, Y. (2009). Crushability of concrete debris: experiments and DEM simulation, Journal of Applied Mechanics, JSCE, 12, 489-496 (in Japanese).

Lockner, D., Byerlee, J. D., Kuksenko, V., Ponomarev, A., & Sidorin, A. (1991). Quasi-static fault

growth and shear fracture energy in granite. Nature, 350(6313), 39-42.

Togo, T., & Shimamoto, T. (2012). Energy partition for grain crushing in quartz gouge during subseismic to seismic fault motion: an experimental study. Journal of Structural Geology, 38, 139-155.

Melosh, H. J. (1989). Impact cratering: A geologic process. Research supported by NASA. New York, Oxford University Press (Oxford Monographs on Geology and Geophysics, No. 11).

Billam, J. (1971). Some aspects of the behaviour of granular materials at hight pressures. Proceedings of the Roscoe memorial symposium, Cambridge, 69-80.

Lee, D. M. (1992). The angles of friction of granular fills. Ph.D. dissertation, University of Cambridge.

Nakata, A. F. L., Hyde, M., Hyodo, H., & Murata. (1999). A probabilistic approach to sand particle crushing in the triaxial test. Geotechnique, 49(5), 567-583.

Weibull, W. (1951). A statistical distribution function of wide applicability. Journal of applied mechanics, 18(3), 293-297.

McDowell, G. R., and M. D. Bolton. (1998). On the micromechanics of crushable aggregates. Geotechnique, 48.5, 667-679.

Zhang, Y. D., Buscarnera, G., & Einav, I. (2015). Grain size dependence of yielding in granular soils interpreted using fracture mechanics, breakage mechanics and Weibull statistics. Géotechnique, 66(2), 149-160.

Stefanou, I., & Sulem, J. (2016). Existence of a threshold for brittle grains crushing strength: two- versus three-parameter Weibull distribution fitting. Granular Matter, 18(2), 1-10.

Nakata, Y., Hyodo, M., Hyde, A. F., Kato, Y., & Murata, H. (2001). Microscopic particle crushing of sand subjected to high pressure one-dimensional compression. Soils and foundations, 41(1), 69-82.

Schofield, A., & Wroth, P. (1968). Critical state soil mechanics. McGraw-Hill.

Pestana, J. M., & Whittle, A. J. (1995). Compression model for cohesionless soils. Géotechnique, 45(4), 611-632.

Jaeger, J. C. (1967). Failure of rocks under tensile conditions. In International Journal of Rock Mechanics and Mining Sciences & Geomechanics Abstracts (Vol. 4, No. 2, pp. 219-227). Pergamon.

Kendall, K. (1978). The impossibility of comminuting small particles by compression. Nature, 272(5655), 710-711.

Turcotte, D. L. (1986). Fractals and fragmentation. Journal of Geophysical Research: Solid Earth, 91(B2), 1921-1926.

Steacy, S. J., & Sammis, C. G. (1991). An automaton for fractal patterns of fragmentation. Nature, 353(6341), 250.

Marks, B., & Einav, I. (2015). A mixture of crushing and segregation: The complexity of grainsize in natural granular flows. Geophysical Research Letters, 42(2), 274-281.

Mandelbrot, B. B. (1982). The fractal geometry of nature. 1982. W.H. Freeman and company, New York.

Oda, M., & Konishi, J. (1974). Microscopic deformation mechanism of granular material in simple shear. Soils and foundations, 14(4), 25-38.

Majmudar, T. S., & Behringer, R. P. (2005). Contact force measurements and stress-induced anisotropy in granular materials. Nature, 435(7045), 1079-1082.

Hartmann, W. K. (1969). Terrestrial, lunar, and interplanetary rock fragmentation. Icarus, 10(2), 201-213.

Matsushima, T., Yamashita, K. & Yamada, Y. (2014). Plastic compression of sands due to grain crushing under high pressure, Proc. COMPSAFE: Computational Engineering and Science for Safety and Environmental Problems, 668-670.

Borkovec, M., De Paris, W., Peikert, R. (1994). The Fractal Dimension of the Apollonian Sphere Packing, Fractals, 2(4), 521–526.

Kitajima, H., Chester, J. S., Chester, F. M., & Shimamoto, T. (2010). High‐speed friction of disaggregated ultracataclasite in rotary shear: Characterization of frictional heating, mechanical behavior, and microstructure evolution. Journal of Geophysical Research: Solid Earth, 115(B8).

Melosh, H.J. (1989). Impact cratering –a geologic process-, Oxford monographs on geology and geophysics, no.11, Oxford University press, 253p.

Tsuchiyama, A., Uesugi, M., Matsushima, T., Michikami, T., Kadono, T., Nakamura, T., Uesugi, K., Nakano, T., Sandford, S.A., Noguchi, R., Matsumoto, T., Matsuno, J., Nagano, T., Imai, Y., Takeuchi, A., Suzuki, Y., Ogami, T., Katagiri, J., Ebihara, M., Ireland, T.R., Kitajima, F., Nagao, K., Naraoka, H., Noguchi, T., Okazaki, R., Yurimoto, H., Zolensky, M.E., Mukai, T., Abe, M., Yada, T., Fujimura, A., Yoshikawa, M., Kawaguchi, J. (2011). Three-Dimensional Structure of Hayabusa Samples: Origin and Evolution of Itokawa Regolith, Science, 333(6046), 1125-1128 (DOI: 10.1126/science.1207807).

Suzuki, A., Matsushima, T. (2014). Meso-scale structural characteristics of clay deposit studied by 2D Discrete Element Method, Proc. IS-Cambridge, Geomechanics from Micro to Macro, Soga et al. Eds, Taylor & Francis, ISBN 978-1-138-02707-7, 33-40.